# Universal Nano-Bead Emitter Inks for Programmable Nanometric Fluorescent Architectures

*Ilya Olevsko, Maria Shehadeh, Dmytro Ohorodniichuk, Leonid Weisman, Rotem Golan, Martin Oheim, Gerardo Byk, and Adi Salomon\**

I. Olevsko, M. Shehadeh, D. Ohorodniichuk, A. Salomon
Chemistry Department, Bar-Ilan University, 529000, Ramat-Gan, Israel
BINA Institute for Nanotechnology and Advanced Materials, Bar-Ilan University, 529000, Ramat-Gan, Israel
E-mail: adi.salomon@biu.ac.il

G. Byk
Chemistry Department, Bar-Ilan University, 529000, Ramat-Gan, Israel

L. Weisman, R. Golan
Claro3d Nano Printing Solutions Ltd., Modiin, Israel

M. Oheim
Université de Paris, CNRS, SPPIN - Saints-Pères Paris Institute for the Neurosciences, F-75006 Paris, France

Funding: This study was financed by a joint Franco-Israeli PHC Maimonide grant (CC-NanoCOQ, to M. Oheim and A. Salomon).

Keywords: Nanometric thin films, Fluorescent nanomaterials , Digital manufacturing, Functional inks , Laser-induced forward transfer (LIFT)

**Abstract**

Fabricating brightly fluorescent layers with nanometric thickness and digitally controlled lateral structuration remains a challenge for next-generation photonic devices, optical calibration standards, and biocompatible interfaces. Here, we introduce Nano-Bead Emitters (NBEs), hydrogel nanoparticles covalently functionalized with fluorophores, as a universal, water-processable ink platform for fabricating programmable nanometric fluorescent architectures. By immobilizing fluorophores within a charged nanohydrogel scaffold, the platform entirely decouples film morphology from dye solubility. This molecule-independent strategy enables spectrally distinct, inherently water-insoluble dyes to be processed using a single, standardized aqueous ink formulation. Combined with laser-induced forward transfer (LIFT) printing, this additive approach yields highly uniform fluorescent layers (~7 nm thickness, sub-nanometric roughness). This structural invariance produces complex multicolor patterns sharing identical thickness and surface morphology across all spectral channels, a critical requirement for quantitative optical calibration. Furthermore, LIFT printing provides programmable, layer-by-layer control over fluorescence intensity via successive deposition cycles, yielding precisely tunable brightness without aggregation-caused quenching. This maskless technique enables rapid, high-fidelity printing of both monochromatic and multicolor patterns over macroscopic areas with absolute spatial resolution. Finally, these universally compatible NBE inks stably deposit onto diverse substrates (glass, polymers, semiconductors, metasurfaces), effectively bridging scalable manufacturing with high-performance integrated photonic systems.

# 1. Introduction

The fabrication of multifunctional thin films with nanometric thickness and controlled lateral structuration is a central challenge in next-generation photonics, biotechnology, and optical security. These applications demand scalable and chemically versatile surface coatings capable of high-density optical encoding, multiplexed sensing, and secure authentication[1–3]. This technological demand has fueled major advances in materials chemistry and nanophotonics, including the on-surface growth of highly oriented molecular frameworks for optoelectronics [4,5], ultrathin metasurfaces for multispectral control and nonlinear holography [6,7] and nanostructured metal oxide thin films for advanced light harvesting. [8] Related requirements arise in optical security and anti-counterfeiting, where complex fluorescent patterns are used as authentication markers, and in applications involving direct contact with biological or consumable products, where biocompatibility and non-toxicity are essential.[9,10]

In parallel, optical metrology increasingly relies on well-defined fluorescent thin films as reference layers for near-interface and evanescent-wave microscopy. Such layers enable nanometric axial calibration and quantitative localization of fluorescent emitters, including in live-cell imaging[11,12]. Building on this approach, we recently introduced multilayer fluorescent architectures that translate axial position along the optical axis into color-encoded signals, providing a direct route to multicolor axial calibration standards. [13]

The successful translation of functional nanomaterials into modern photonic devices relies critically on the development of scalable, economical, and practical fabrication technologies.[14] However, Existing fabrication techniques for fluorescent thin films involve fundamental trade-offs between uniformity, scalability, and patterning flexibility. Spin coating remains the laboratory standard for producing high-quality ultra-thin films, but it is inherently restricted to uniform, monochromatic layers and suffers from substantial material loss, often exceeding 90% [15,16]. Soft lithography and microcontact printing provide low-cost access to nanometric patterning, yet rely on static physical masters that limit rapid design iteration and complicate multicolor overlay.[17,18] In contrast, top-down lithographic approaches offer high spatial resolution but require harsh processing conditions, including high vacuum and ultraviolet exposure, which are frequently incompatible with fluorescent dyes and biologically derived materials.[19]

Digital printing technologies, such as inkjet and electrohydrodynamic jet printing, offer a potential route toward digitally programmable deposition through drop-on-demand patterning.[20,21] However, inkjet printing of molecular dye solutions is often limited by the coffee-ring effect, in which evaporation-induced capillary flows drive solutes toward the droplet perimeter, producing non-uniform thickness and concentration profiles.[22,23] These effects compromise optical flatness and hinder the fabrication of homogeneous fluorescent reference layers. Moreover, achieving reliable control over film thickness in the sub-100 nm regime remains challenging, as increasing dye concentration frequently promotes aggregation and fluorescence quenching in conventional molecular inks.

Here, we introduce Nano-Bead Emitters (NBEs) as a universal, water-processable ink for digital fabrication of ultra-thin fluorescent films and patterned architectures. NBEs are modular hydrogel nanoparticles that covalently attached to fluorophores within a charged, hydrated polymer matrix, building on established biocompatible hydrogel chemistries.[24] This design decouples film formation from the solubility and chemical identity of the dye, enabling spectrally diverse emitters, including dyes that are not intrinsically water-soluble, to be processed using a single aqueous formulation.[25]

The NBE ink exhibits intrinsic auto-smoothing behavior arising from electrostatic repulsion between negatively charged nanoparticles and hydrogel-mediated steric stabilization. These interactions suppress aggregation and vertical stacking while promoting uniform lateral spreading, yielding ultra-thin (~7 nm), morphologically homogeneous films that are largely independent of fluorophore identity and labeling density.

Combined with laser-induced forward transfer (LIFT) printing[26–28], this approach enables digitally programmable fluorescent architectures across length scales that are not readily accessible using conventional deposition methods.[29,30] Here, we demonstrate uniform nanometric films, multicolor patterning with invariant thickness across spectral channels, and programmable control over emission intensity and interfacial mixing.

## 2. Results and discussion

### 2.1. Universal Morphology and Decoupling of Structural Properties

**Figure 1** demonstrates that the NBE platform decouples the optical properties of the fluorophore from the structural characteristics of the deposited layer, consistent with our previous findings. [25]

Three chemically distinct dyes- ATTO 425 (blue emission), ATTO 488 (green emission ), and ATTO 490LS (red emission)-were covalently conjugated to the NBE scaffold. The NBE have been printed using the proprietary and patented LIFT printing technology developed by Claro 3D Nano Printing Solution Ltd. (refers hereinafter as "Claro 3D"). Importantly, all NBE inks were processed from the same aqueous solvent system, despite the different intrinsic solubilities of the corresponding free dyes. Independent of fluorophore identity, the resulting films exhibit indistinguishable morphology and lateral uniformity over millimeter-scale areas (Figure 1A , Figure S1).

This observation confirms that thin-film assembly is governed by the hydrogel nanoparticle carrier and the printing process, rather than by fluorophore chemistry or solvent-specific effects.

Fluorescence emission spectra of the printed layers closely resemble those of the corresponding free dyes in solution (Figure 1B). The only exception is NBE(ATTO 490LS), which displays a reproducible blue shift, consistent with previously reported behavior upon incorporation into the NBE matrix.[25]

The formation of uniform, ultra-thin NBE films is governed by a combination of electrostatic and molecular-level interactions. The negatively charged hydrogel nanoparticles experience a lateral electrostatic repulsion, which suppresses agglomeration and promotes homogeneous spreading upon deposition. To stabilize this dispersed layer, a positively charged primer interface is introduced. Coating glass or fused-silica substrates with a nanometric layer of poly(diallyldimethylammonium chloride) (PDDA) provides electrostatic attraction that anchors the NBEs to the surface, yielding a stable and laterally uniform coating.

In parallel, fluorophores are covalently attached to the hydrogel scaffold through a limited number of amine ligands, resulting in an average intermolecular separation of several nanometers (~5 nm). This controlled spacing prevents close-contact interactions between neighboring emitters, suppressing π–π stacking and thereby avoiding aggregation-induced quenching. Together, these electrostatic and molecular design features ensure reproducible film morphology and fluorescence performance, establishing the NBE platform as a generalizable printing vehicle for chemically diverse emitters.

AFM further confirmed a flat topography with an average thickness of approximately 7 nm, an average roughness ($R_a$) of 0.96 nm, and maximum height deviations below 20 nm (Figure S2). This surface quality extends beyond the nanometric scale to macroscopic dimensions. Together, these parameters make printed NBE layers reliable intensity and focal standards for color-multiplexed or height-encoded calibration tools.[13,25]

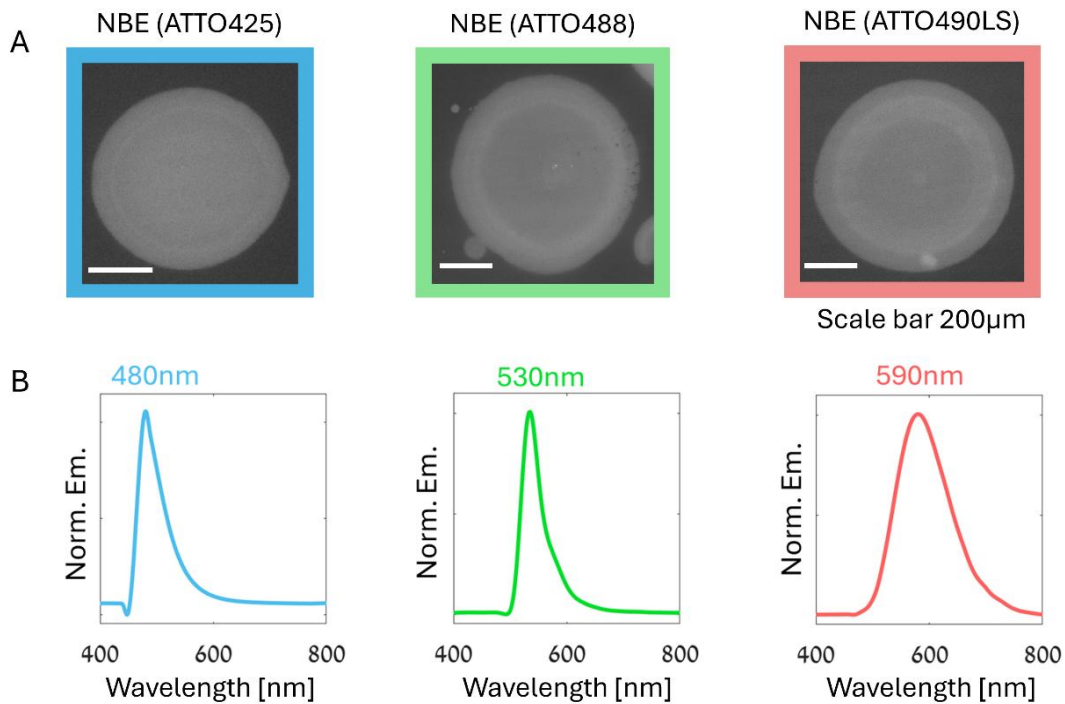

**Figure 1. Homogeneous behavior of multicolor NBE thin films.** A) Epi-fluorescence images of Claro 3D's LIFT-printed NBE spots using different fluorophores (from left to right): NBE(ATTO 425), NBE(ATTO 488), and NBE(ATTO 490LS), respectively. Each image displays a representative spot from a larger array (100 spots printed in a single cycle), demonstrating that the layer morphology and pattern remain consistent across different colors. Scale bar: 200 µm B) Corresponding emission spectra of the resulting layers. Images and spectra were recorded upon 405 nm excitation with a 450 nm long-pass (LP) filter for NBE(ATTO 425), and 488 nm excitation with a 500 nm LP filter for NBE(ATTO 488) and NBE(ATTO 490LS), respectively.

## 2.2 Control over NBE Patterning and Morphology via Surface Chemistry and Printing Parameters

Claro 3D's proprietary Laser-Induced Forward Transfer (LIFT) [28] enables rapid printing of NBE, with a unique capability to deposit highly diluted, low-viscosity, water-based materials that are challenging or not feasible to print using conventional technologies. Claro 3D's LIFT based printer print patterns with micrometer-scale precision, allowing arrays comprising more than 100 features to be generated within seconds. Pattern morphology is governed by substrate surface charge together with a limited set of digital printing parameters, including NBE concentration, array pitch (p), pattern symmetry, and nozzle orifice size.

As summarized in **Figure 2**, systematic variation of these parameters provides deterministic access to a broad range of morphologies, from isolated spots to dense cubic and hexagonal arrays and continuous layers. In the absence of a positively charged PDDA primer layer, NBE inks exhibit poor wetting and thus limited spreading, resulting in thick, non-uniform features. In contrast, PDDA-functionalized substrates promote efficient wetting and lateral spreading, enabling the formation of homogeneous nanometric films (Figure S3). Continuous layers are obtained by combining elevated NBE concentration with multiple printing cycles and controlled spatial offsets between successive arrays.

Finally, the generality of the NBE-LIFT approach was demonstrated across multiple substrates, including plasmonic Au/SiO$_x$ films, silicon with native oxide, and flexible polymer layers (Figure S4). While spreading dynamics depend on surface adhesion and electrostatic interactions, stable and well-defined patterns were obtained in all cases, including on Au/SiO$_x$ substrates with PDDA functionalization.

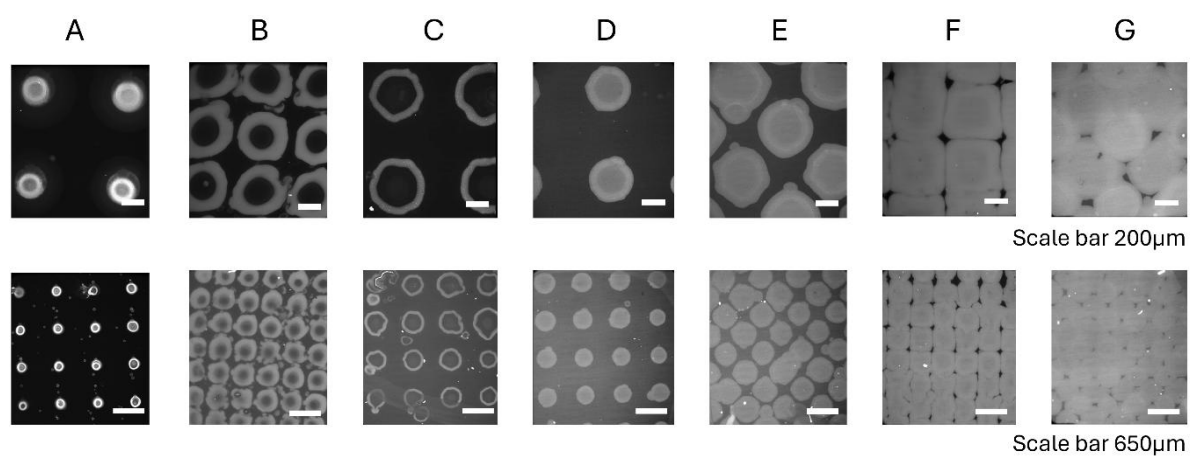

**Figure 2. Printing parameters control NBE pattern morphology.** Representative fluorescence micrographs of Claro 3D's LIFT-printed arrays of NBE(ATTO490LS), upon 488-nm excitation. Similar patterns were obtained for other NBE variants. Top, high-magnification images (10× objective, scale bar: 200 μm), and low-mag view, bottom, (4× objective, scale bar: 650 μm). Patterns was controlled by modulating the NBE concentration, printing pitch (p), symmetry, and nozzle orifice size. A) 100% concentration printed on a substrate without a PDDA primer layer (p = 800 μm, cubic array, 80 μm orifice). B) 1% concentration (p = 500 μm, cubic, 150 μm orifice). C) 2% concentration (p = 800 μm, cubic, 80 μm orifice). D) 3% concentration (p = 800 μm, cubic, 80 μm orifice). E) 6% concentration (p = 800 μm, hexagonal, 80 μm orifice). F) 6% concentration (p = 500 μm, cubic, 150 μm

orifice). G) 6% concentration (p = 600 μm, hexagonal, 80 μm orifice, 2 cycles with offset cubic arrays).

## 2.3 Fluorescence Intensity Control via Multiple Printing Cycles

NBE–LIFT printing enables quantitative control of fluorescence intensity through repeated deposition at identical coordinates. Successive printing cycles increase the local surface density of NBE particles, providing discrete tuning of emission intensity. This behavior was investigated using NBE (ATTO 425) under conditions where coffee-ring formation is suppressed and the PDDA primer remains below saturation (Figure S5).

Fluorescence intensity scales with the number of printing cycles. Doubling the number of depositions passes yields an approximately twofold increase in emission intensity for both continuous films (**Figure 3**A,B) and discrete spot arrays (Figure 3C,D), while preserving macroscopic film uniformity over millimeter-scale areas (Figure S1).

Repeated depositions also affect morphology. Additional printing cycles lead to an increase in spot diameter and a moderate reduction in internal homogeneity (Figure S5). Thus, multi-cycle printing enables controlled intensity enhancement with a defined trade-off in feature size and surface roughness.

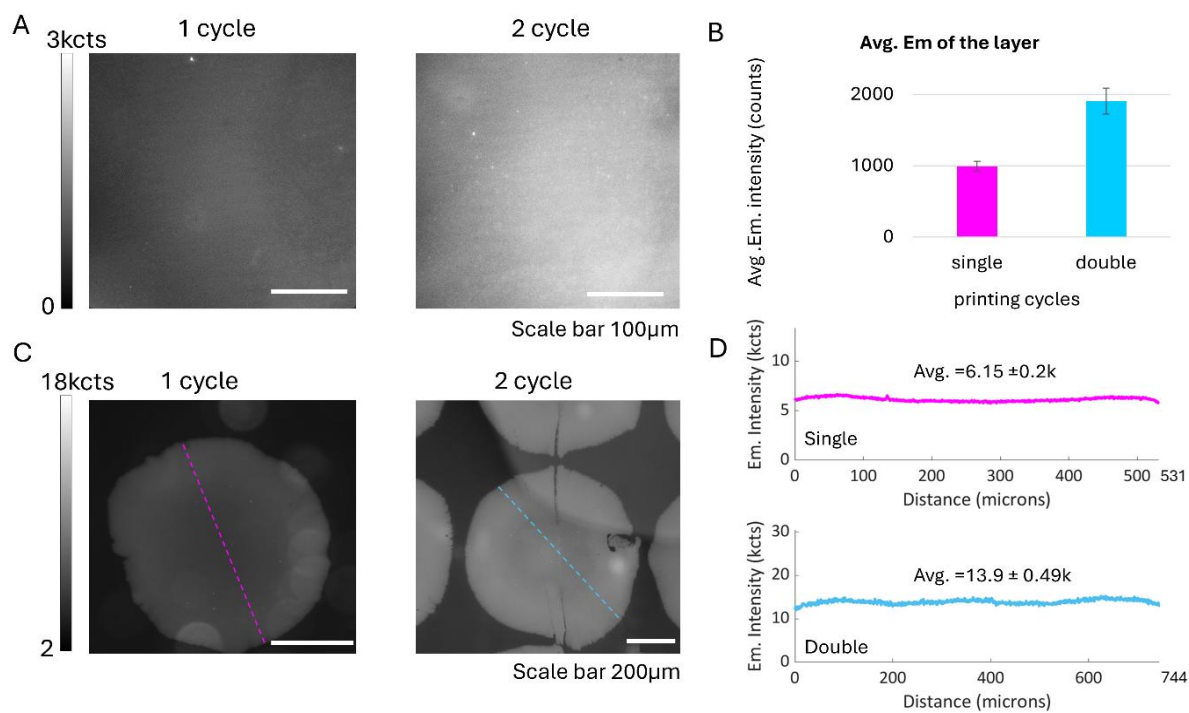

**Figure 3. Control of fluorescence intensity via multiple printing cycles.**

NBEs labeled with ATTO 425 (excitation: 405 nm) demonstrate tunable emission intensity through repeated deposition. (A) Fluorescence images of continuous layers of one (left) and two (right) cycles. (B) Mean of fluorescence intensities extracted from central regions. Doubling the number of printing cycles yields an approximately twofold increase in signal (single: ~$9.9\times10^2$ counts; double: ~$1.9\times10^3$ counts). Error bars represent standard deviation. (C) Fluorescence micrographs of discrete spots (p = 800 μm, cubic array) printed with one (left) and two (right) cycles. (D) Representative intensity profiles along the lines indicated in (C), confirming proportional scaling of peak intensity with the number of printing

## 2.4 Multi-Color Fabrication and Controllable Dye Mixing

Claro 3D's LIFT printing enables spatially resolved deposition of multiple NBE inks on a single substrate. Sequential printing of NBEs labeled with ATTO 488 (green), ATTO 425 (blue), and ATTO 490LS (red) yields complex multicolor architectures with micrometer-scale positional accuracy (Figure 4). Concentric structures are obtained by depositing a second dye onto a pre-printed spot (**Figure 4**A), while adjacent fluorophore domains can be patterned side-by-side without detectable spectral crosstalk (Figure 4B,C).

Control over color mixing is achieved through the inter-deposition delay, which determines the drying state and mobility of the underlying NBE layer. As shown in Figure S6, short delays (~1 min) preserve residual solvent and particle mobility. During the second deposition, electrostatic repulsion between negatively charged NBE populations displaces the mobile particles of the first layer, producing sharply segregated interfaces. Longer delays (>2 min) immobilize the first layer on the substrate; subsequent deposition then forms an overlay region where both fluorophore populations coexist, resulting in controlled interfacial mixing.

Cross-sectional intensity profiles confirm minimal spectral overlap in the short-delay regime, indicating nanometric-scale separation between color domains (Figure S7). In contrast, simultaneous deposition of identical inks produces larger micron-scale gaps due to stronger electrostatic repulsion between fluid droplets (Figure S7C). These results demonstrate deterministic control over spectral segregation and interfacial mixing using a single temporal printing parameter.

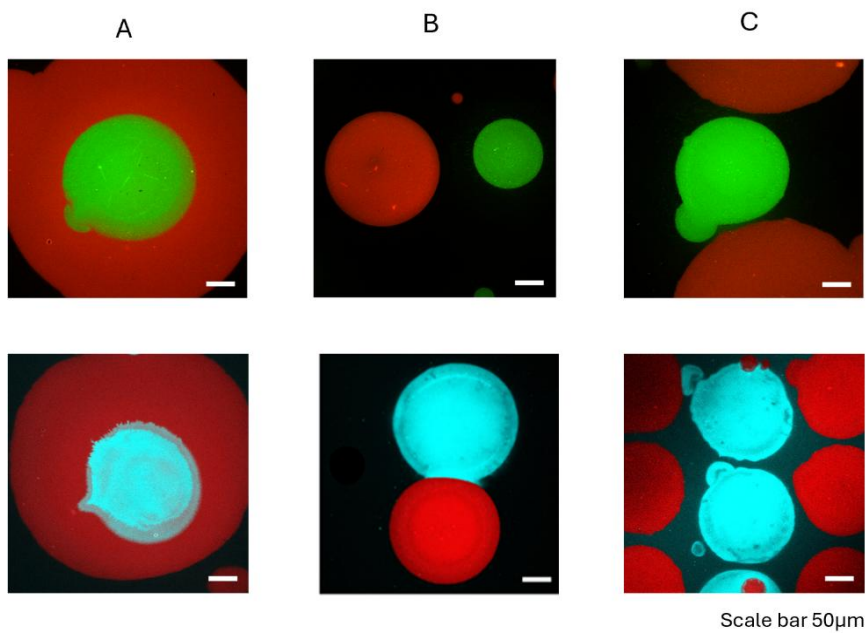

Scale bar 50μm

**Figure 4. Fabrication of multi-color NBE patterns and control over dye mixing.** Confocal fluorescence images (20×/0.8 NA) of multi-color NBE patterns printed on a single optical glass coverslip. NBE(ATTO 488) is displayed in pseudo color green, NBE(ATTO 425) in blue, and NBE(ATTO 490LS) in red. A) Concentric patterns formed by depositing the second dye (50 μm orifice) onto a pre-printed red spot (150 μm orifice). B, C) Spatially separated patterns with varying inter-spot distances, demonstrating positional precision (Red/Blue: 150 μm orifice; Green: 50 μm orifice).

## 3. Conclusion

We establish NBE as a universal, water-processable platform for the fabrication of ultra-thin fluorescent layers and patterned architectures. By confining fluorophores within a charged hydrogel nanoparticle scaffold, film formation is governed by the carrier rather than the dye, rendering layer morphology; thickness, roughness, and spreading largely independent of fluorophore identity and solubility.[25] This decoupling overcomes a central limitation of conventional molecular inks and enables the processing of dyes that are not intrinsically water-soluble within a single aqueous formulation.

Combined with LIFT printing [26–28], the platform enables deterministic control over pattern morphology, multicolor organization, and emission intensity. Multicolor arrays with invariant thickness across spectral channels are readily produced, while repeated deposition provides discrete intensity tuning without concentration quenching. Control over inter-deposition timing further allows switching between segregated and overlapping dye interfaces.

The approach is compatible with diverse substrates, including glass, silicon, plasmonic Au/SiO$_x$ films, and flexible polymers. Together, these capabilities define a scalable and material-independent strategy for constructing fluorescent architectures with nanometric precision, with immediate implications for optical metrology, multiplexed sensing, and integrated photonic systems.

## 4. Experimental Section

*Synthesis of Nano-Bead Emitters (NBEs):* The synthesis of NBEs relied on the chemical attachment of ATTO 425, ATTO 488, and ATTO 490LS to nanohydrogels (NHGs) via amide bond formation. The complete polymerization and crosslinking protocols were executed as previously reported.[25] Zeta potential characterization of the resulting NHGs yielded a mean value of -6.9 mV, confirming a net negative surface charge. The estimated spatial separation between potential dye binding sites was determined to be roughly 4-5 nm. Because the dimensions of the conjugated molecular emitters are typically 1-2 nm, this inter-fluorophore distance significantly exceeds the characteristic π-π stacking radius, thereby minimizing aggregation-caused fluorescence quenching. Further photophysical and structural properties regarding surface charge and molecular isolation have been detailed previously.[25]

*Substrate Preparation and Maskless LIFT Printing:* Glass coverslips (170 μm thickness; ISOLab, 24 x 50 mm, or Marienfeld Superior, 25 mm diameter) were utilized as optical substrates. Cleaning was performed by sonication in Hellmanex III solution, followed by sequential washing cycles with 99% ethanol and deionized water (DIW). To facilitate adhesion, a primer layer was deposited via a layer-by-layer (LbL) approach utilizing the positively charged polymer poly(diallyldimethylammonium chloride) (PDADMAC; average Mw 400000-500000, Sigma-Aldrich, CAS 26062-79-3). A 20 wt% aqueous stock solution was diluted 1:10 with DIW to yield a final concentration of 2 wt%. Subsequently, 20 μL of this PDADMAC solution was drop-cast onto the cleaned substrates. Following a 60 s retention period, excess solution was removed via a stream of compressed air, and the functionalized substrates were maintained in a dry desiccator prior to the printing phase.

The printing inks comprised the aqueous NBE dispersion. Post-dialysis solutions, originally loaded at 3 mg mL$^{-1}$, underwent slight osmotic dilution, yielding a final "100%" stock concentration estimated between 2-3 mg mL$^{-1}$. Subsequent concentrations utilized for patterning (1%-25%) were prepared via volumetric dilution of this master stock.

Deposition was executed using a proprietary maskless Laser-Induced Forward Transfer (LIFT) platform (Claro 3D, patented technology [26–28]). This specific architecture was selected for its capability to reliably deposit highly diluted, low-viscosity, water-based materials at high resolutions (below 20 μm), independent of surface topography or shear force sensitivity. The system utilizes a microfluidic chip to simultaneously handle up to four distinct aqueous inks, employing interchangeable nozzle orifices of 120, 80, and 50 μm. Droplet ejection was driven by a focused, short-pulse laser that generates a transient plasma, propelling droplets from the microfluidic orifice. Optimal deposition was achieved employing 87% laser power at a pulse frequency of 6 Hz. A motor-driven printhead facilitated precise spatial translation across the substrate to generate defined arrays, allowing droplets to dry in situ. Each printing cycle deposited a 10 x 10 array (100 spots), with adjustable inter-cycle delays employed to govern drying kinetics and final lattice geometry.

*Optical Characterization and Fluorescence Microscopy:* Emission spectra and epi-fluorescence imaging were acquired utilizing an inverted microscope system (IX83, Olympus) interfaced with an Iso-Plane SCT320 spectrophotometer (600 nm blaze, 50 grooves mm$^{-1}$ grating) and a PIXIS 1024 eXcelon charge-coupled device (CCD) camera (Teledyne Princeton Instruments). Excitation was provided by an Excelitas X-Cite 120Q illumination unit coupled with appropriate dichroic mirrors. Spectral data (Figure 1) were collected via a 200 μm slit aperture; NBE-ATTO425 was excited at 405 nm (emission collected via a 450 nm long-pass filter), while NBE-ATTO488 and NBE-ATTO490LS were excited at 488 nm (emission collected via a 500 nm long-pass filter). Spectrometer calibration was performed utilizing a Princeton Instruments IntelliCal source. Emission spectra represent an average across 1023 pixel rows, processed via MATLAB software. For epi-fluorescence mapping (Figures 1, 2, 3A, S3-S5), the spectrometer grating was bypassed with a mirror, and images were acquired through 4x (NA 0.15) and 10x (NA 0.25) objectives with exposure times ranging from 100 to 500 ms. Quantitative intensity analysis was conducted in MATLAB by averaging signals across a central 500 x 500 pixel region of interest (ROI); all comparisons were standardized to samples exhibiting equal optical absorbance at the excitation wavelength. Error bars represent the standard deviation.

Several fluorescent images (Figures 3B, S1) were executed using an Apotome 3 system (Carl Zeiss) equipped with an Axiocam 807 mono camera, an HXP 120 V light source, and a standard DAPI module. Intensity profiles were calculated as raw counts versus lateral distance along the indicated scan lines.

Confocal laser scanning microscopy (Figures 4, S8, S9) was performed on an LSM 710 system mounted on an Axio Imager 2 stand (Carl Zeiss) employing a Plan-Apochromat 20x/0.8 NA air objective. Micrographs were captured at 1024 x 1024 pixel resolution (12-bit depth, 0.415 μm pixel size) with an 8-line average and a 0.64 μs pixel dwell time. Dual-channel architectures were interrogated using distinct excitation/emission pairings: Red-Blue samples (excitation: 405 nm; emission: 420-486 nm and 578-656 nm) and Green-Red samples (excitation: 488 nm; emission: 480-539 nm and 607-705 nm). Post-acquisition composite overlays and channel-independent intensity cross-sections (Figure S9) were generated utilizing MATLAB.

*Surface Topography and Atomic Force Microscopy (AFM):* The average surface roughness (Ra) of the printed thin films (Figure S2) was evaluated using a Bio FastScan Atomic Force Microscope (Bruker AXS). Scanning was conducted in soft tapping mode employing a silicon probe with a silicon nitride cantilever (FASTSCAN-B, Bruker; tip thickness: 0.3 μm, length: 30 μm, nominal spring constant: 1.8 N m$^{-1}$, resonance frequency: 450 kHz). Topographical maps were captured at a resolution of 512 points per line. Raw AFM data were analyzed and leveled utilizing a 2nd-degree polynomial flattening algorithm within Gwyddion software.

**Data Availability Statement**

All relevant data supporting the findings of this study are available within the article, its

Supplementary Information, and from the corresponding author upon reasonable request.

**Supporting Information**

Supporting Information is available from the Wiley Online Library or from the author.

Supporting Information

**Universal Nano-Bead Emitter Inks for Programmable Nanometric Fluorescent Architectures**

*Ilya Olevsko, Maria Shehadeh, Dmytro Ohorodniichuk, Leonid Weisman, Rotem Golan, Martin Oheim, Gerardo Byk, and Adi Salomon\**

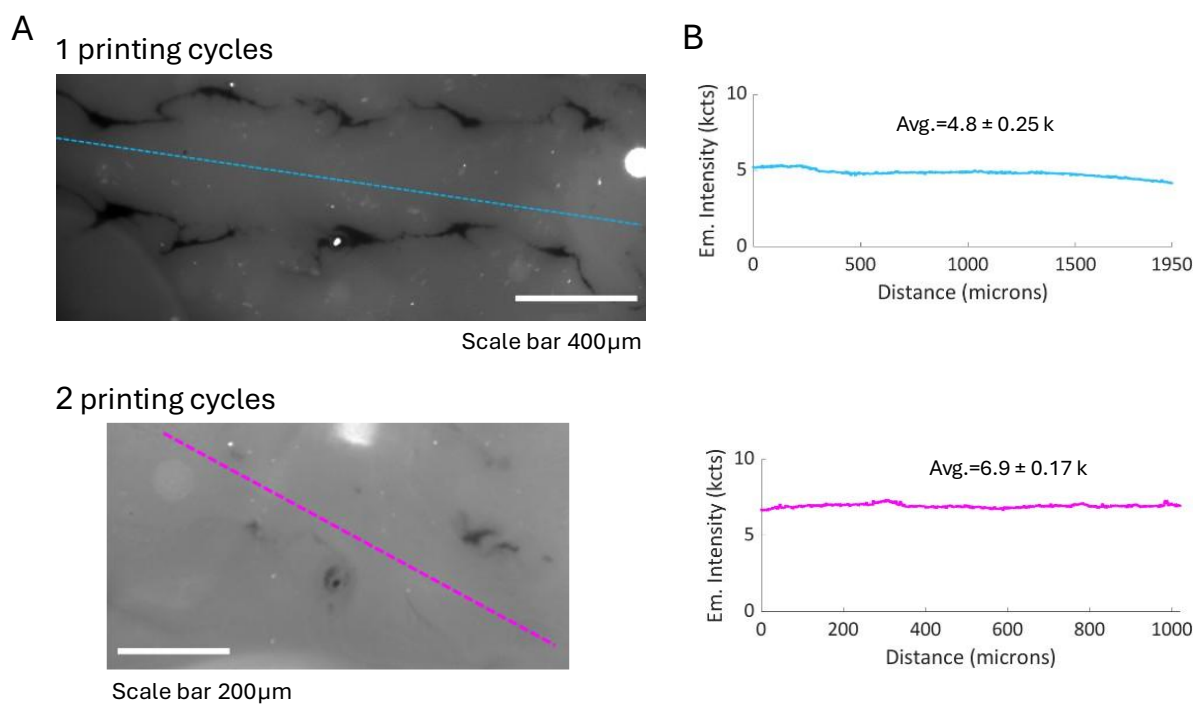

**Figure S1: Long-range uniformity of NBE layers.** Fluorescence characterization of NBE (ATTO 425) layers printed in a cubic array ($p = 400$ μm) with 1 or 2 printing cycles, imaged using a Zeiss Apotome 3 setup. A) Fluorescence images (excitation: 405 nm) covering a macroscopic field of view. B) Intensity cross-sections taken along the dashed lines indicated in (A), with curve colors corresponding to the specific lines in the images.

The profiles demonstrate that layer homogeneity is preserved over millimeter-scale distances. Furthermore, the additional printing session significantly increases fluorescence intensity without disrupting the uniformity of the film. As expected, the standard deviation (SD) is slightly higher for the double-printed layer; however, the SD remains satisfyingly low even though the measured length is twice as large, further validating the robust uniformity of the layers.

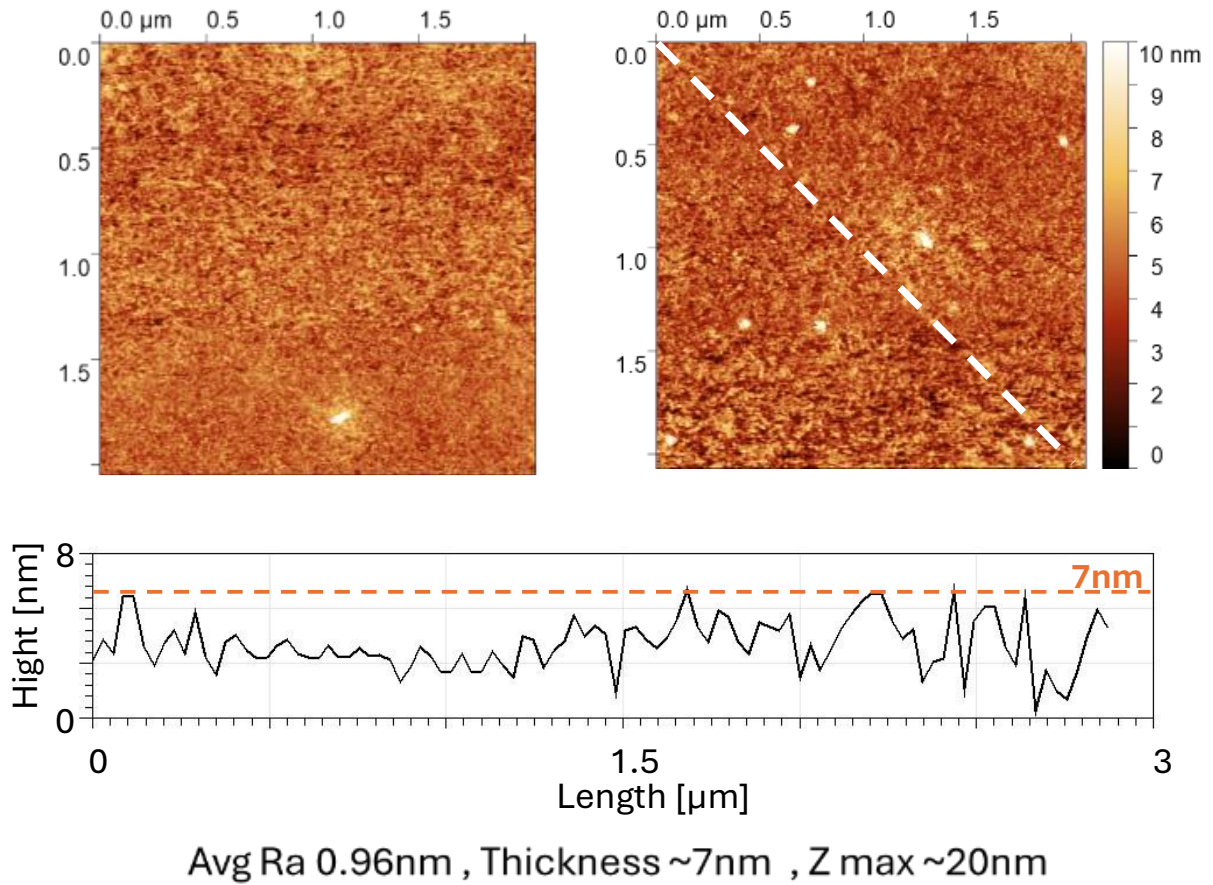

**Figure S2: Thickness and roughness characterization of NBE layers.** Atomic Force Microscopy (AFM) analysis (soft tapping mode, 2x2μm) of NBE spots printed on a BK7 glass substrate pre-treated with 2% PDDA. Topographic scans of two representative samples reveal a smooth and uniform surface morphology, characterized by an average roughness (*Ra*) of 0.96 nm and maximum height deviations (*Z*) below 20 nm. A representative Z-axis cross-section demonstrates an average layer thickness of approximately 7 nm. This behavior is consistent with expected results for NBEs (I. Olevsko et al., *Adv. Optical Mater.* 2025).

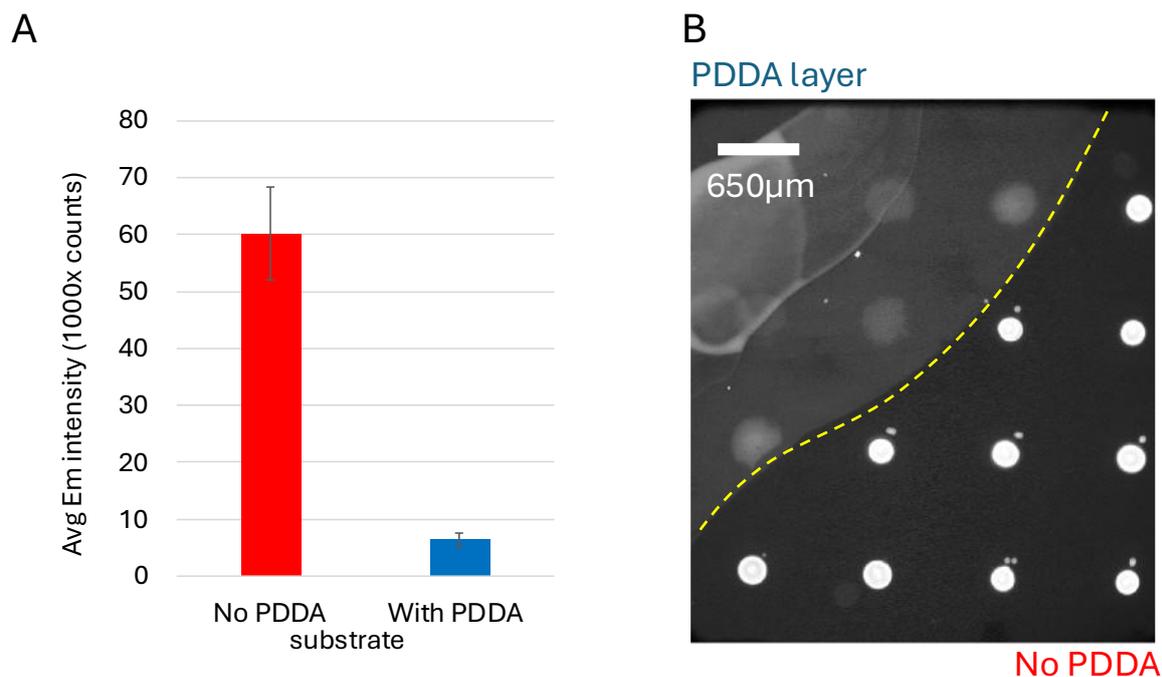

**Figure S3: Impact of PDDA primer on NBE layer fluorescence intensity**. Analysis of an NBE (ATTO 425) layer printed across the boundary of a PDDA primer coating, allowing simultaneous comparison of spots on the PDDA-coated surface and the bare glass substrate. A) Average integrated fluorescence intensity of the printed spots. Blue bars correspond to spots on the PDDA layer, while red bars correspond to spots printed directly on the glass. The spots without PDDA exhibit significantly higher emission intensity, attributed to a much higher density of emitters without quenching -a characteristic feature of the NBE platform. However, the larger error bars indicate lower homogeneity in the absence of the PDDA layer. B) Representative fluorescence image (excitation: 405 nm) captured using an Olympus IX83 setup. The scale bar for this image is 650μm. The image displays the interface of the PDDA layer (marked by a yellow dashed line), showing the transition between the coated and uncoated regions.

At this high concentration, NBEs tend to stack at the center rather than spreading uniformly. Additional printing cycles predominantly increase central accumulation without significantly enhancing the brightness of the surrounding thin film, indicating a limitation in particle spreading. As seen in the low-DR images and cross-sections, the outer layers extend toward adjacent spots (pitch $p$ = 800 μm) but do not fully merge. A distinct dark gap is preserved even after multiple printing sessions, with average gap widths measuring 5.53 μm after 2 cycles and 4.4 μm after 3 cycles (averaged over 3 areas).

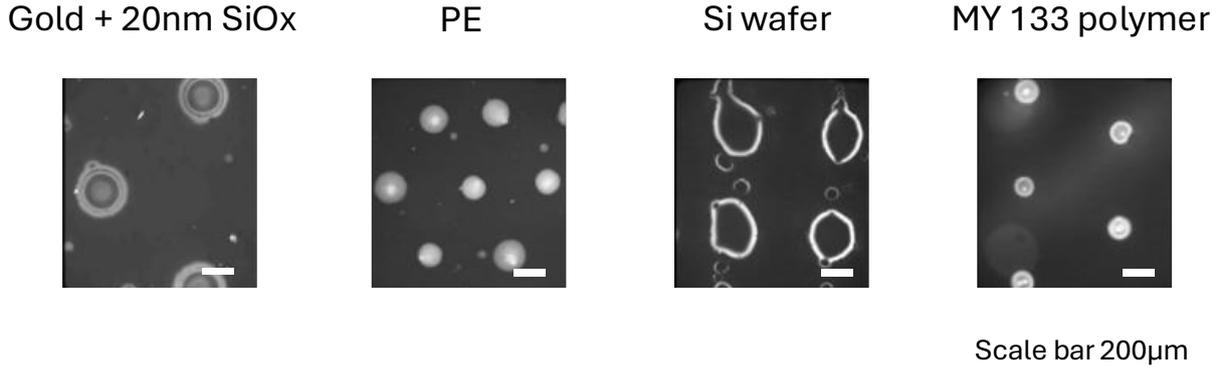

Scale bar 200μm

**Figure S4: Versatility of NBE printing on various substrates.** Epi-fluorescence images (10x objective) demonstrating the adaptability of the NBE platform to four distinct substrate materials with varying surface chemistries. A PDDA primer layer was applied to all substrates prior to printing. Scale bar: 200 μm for all panels. Column 1: Gold (300 nm) covered with 20 nm SiOx. Printed with NBE(ATTO 490LS) (3%, 150 μm orifice, pitch $p$ = 900 μm). Column 2: Corona-treated Polyethylene (PE). Printed with NBE(ATTO 490LS) (3%, 150 μm orifice, pitch $p$ = 500 μm). Column 3: Silicon wafer with native oxide. Printed with NBE(ATTO 488) (2%, 50 μm orifice, pitch $p$ = 950 μm). Column 4: MY-133 polymer layer (~200 nm, CT treated). Printed with NBE(ATTO 488) (3%, 50 μm orifice, pitch $p$ = 600 μm).

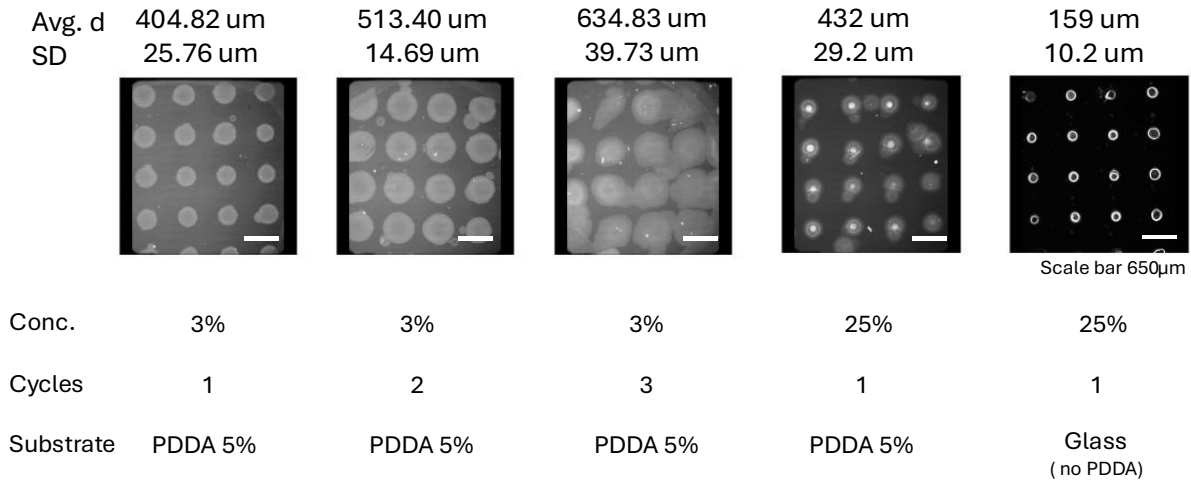

| | | | | | |
|---|---|---|---|---|---|
| Avg. d | 404.82 um | 513.40 um | 634.83 um | 432 um | 159 um |
| SD | 25.76 um | 14.69 um | 39.73 um | 29.2 um | 10.2 um |
| Conc. | 3% | 3% | 3% | 25% | 25% |
| Cycles | 1 | 2 | 3 | 1 | 1 |
| Substrate | PDDA 5% | PDDA 5% | PDDA 5% | PDDA 5% | Glass (no PDDA) |

Scale bar 650μm

**Figure S5: Characterization of NBE spot size and morphology.** Analysis of NBE (ATTO 490LS) spots printed on glass substrates with a 5% PDDA primer layer, as well as on bare glass (no PDDA) for comparison. Printing parameters were maintained at $p$ = 800 μm, orifice 80 μm, energy 92%, and frequency 10 Hz, with a 1-minute delay between cycles. The figure presents the average spot diameter (Avg. $d$) and standard deviation (SD) as a function of concentration and cycle number. The analysis focuses on concentrations ≥ 3%, as lower concentrations resulted in ring-shaped "coffee-ring" structures. Representative wide-field epi-fluorescence images (excitation: 488 nm, 4× objective) are displayed. are displayed. All images are presented with 650μm scale The presented Avg. $d$ values were calculated from multiple fields of view and are consistent with results obtained for other NBE variants.

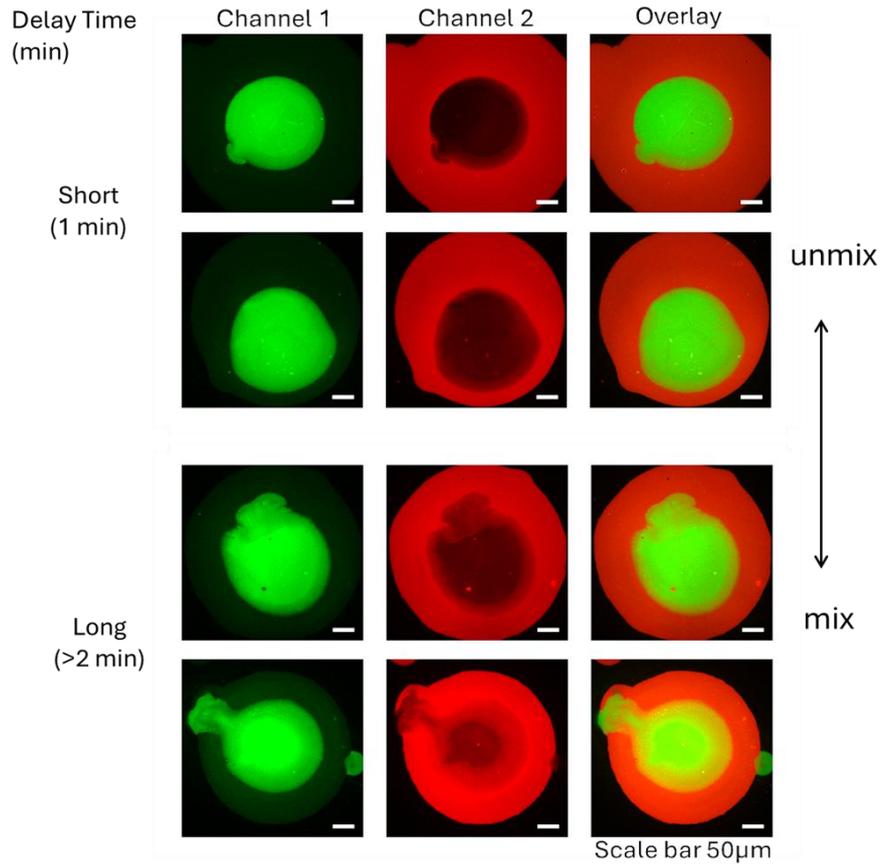
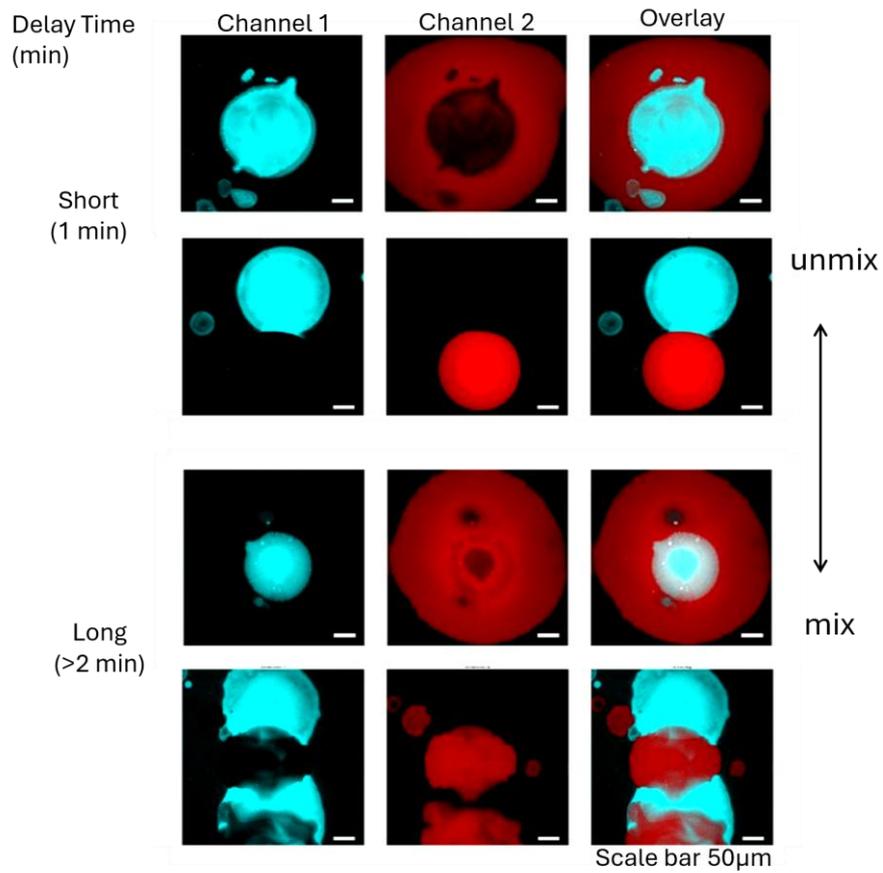

**Figure S6: Interface formation and mixing control via deposition delay.** Confocal fluorescence images characterizing the boundary interaction between sequentially printed NBE layers. NBE(ATTO 490LS) (red) was deposited first, followed by either NBE(ATTO 425) (blue) or NBE(ATTO 488) (green). Each panel displays the individual spectral channels alongside the composite overlay, labeled with the specific delay time applied between printing sessions. **Imaging parameters: Red-Blue (RB):** Excitation: 405 nm; Detection: 420–486 nm (Blue) and 578–656 nm (Red). **Green-Red (RG):** Excitation: 488 nm; Detection: 480–539 nm (Green) and 607–705 nm (Red).

**Mechanism of Interface Formation:** The interface morphology is governed by the hydration state and mobility of the underlying layer at the moment of the second deposition.
**Short Delay (1 min):** Residual solvent retains the mobility of the first NBE layer. Consequently, the electrostatic repulsion from the newly deposited second drop effectively "pushes" the previous NBEs away. This displacement prevents overlap, resulting in a sharp, highly segregated interface.
**Long Delay (>2 min):** The first layer is fully dried and structurally pinned to the substrate. The second deposition cannot mechanically displace the immobilized NBEs, leading to the formation of a co-existence zone at the borders where the two fluorophore populations overlap (observable in the separate spectral channels).

A

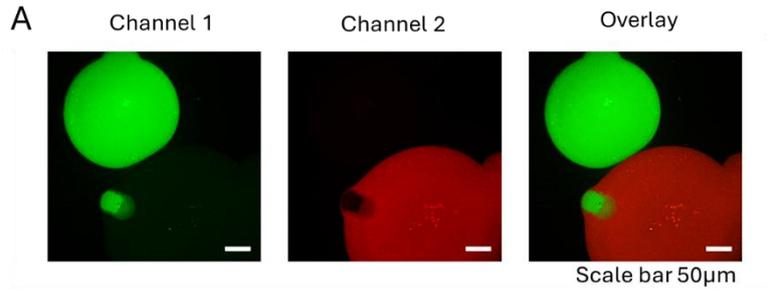

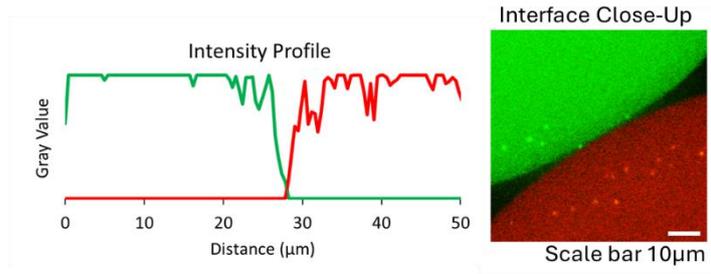

B

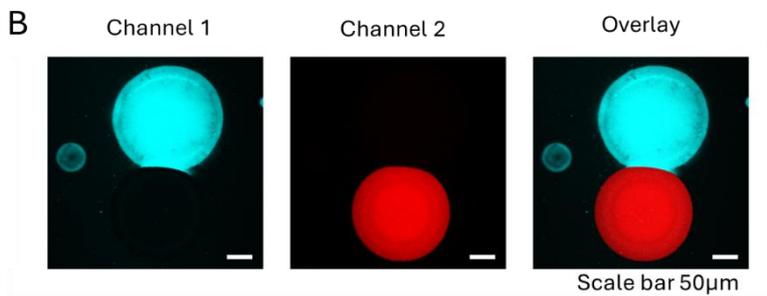

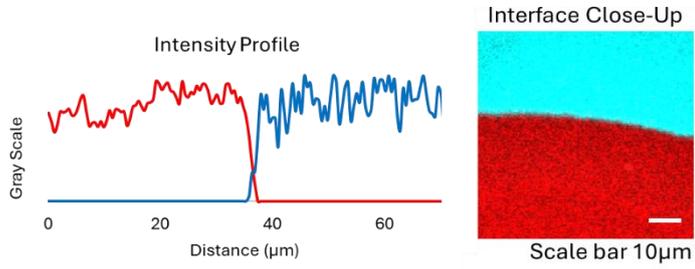

C

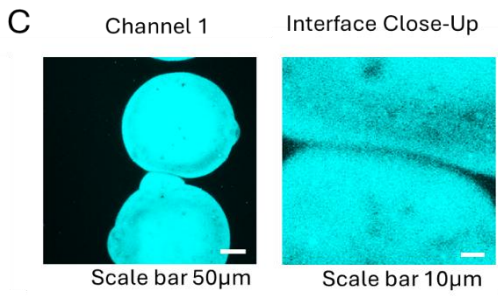

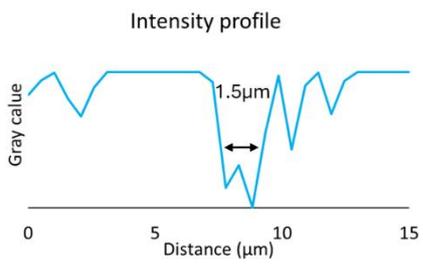

**Figure S7: Characterization of NBE layer interfaces.** Confocal fluorescence analysis of interfaces formed between distinct NBE layers deposited adjacently without intentional spatial overlap. Each panel displays separate spectral channel images, a composite overlay, a high-magnification close-up of the interface region, and an intensity profile plot across the boundary (curve colors correspond to the image channels).
**A)** Interface between NBE(ATTO 490LS) (Red) and NBE(ATTO 488) (Green). The red layer was deposited 1 minute prior to the green layer. Detection channels: 480-539 nm (Green) and 607-705 nm (Red). **B)** Interface between NBE(ATTO 490LS) (Red) and NBE(ATTO 425) (Blue). The red layer was deposited 1 minute prior to the blue layer. Detection channels: 578-656 nm (Blue) and 607-705 nm (Red). **C)** Interface formed between two adjacent NBE(ATTO 425) (Blue) layers deposited simultaneously in a single printing cycle.
The intensity profiles and close-up views reveal the nature of the boundaries formed under these specific deposition conditions. Specifically, they demonstrate that in the absence of intentional spatial overlap, the layers do not tend to diffuse into one another, resulting in a distinct, narrow interface with minimal to no intermixing.